\documentclass{jpaper}

\usepackage[nocompress]{cite}
\usepackage{algorithmic}
\usepackage{array}

\makeatletter
\let\MYcaption\@makecaption
\makeatother

\usepackage[font=footnotesize]{subcaption}

\makeatletter
\let\@makecaption\MYcaption
\makeatother

\usepackage{fixltx2e}
\usepackage{dblfloatfix}
\usepackage[nolessnomore, italic]{mathastext}
\usepackage[T1]{fontenc}
\usepackage[usenames,dvipsnames,svgnames,table]{xcolor}
\usepackage[normalem]{ulem}
\usepackage{enumitem}
\usepackage{setspace}
\usepackage{indentfirst}
\usepackage{footmisc}
\usepackage{fancyhdr}
\usepackage{authblk}
\usepackage{url}
\usepackage[binary-units=true]{siunitx}
\usepackage[us,12hr]{datetime}
\usepackage[keeplastbox]{flushend}
\usepackage[hidelinks]{hyperref}

\newif\ifcameraready
\camerareadytrue

\newcommand{\versionnum}[0]{7}

\ifcameraready
\else
\fi

\ifcameraready
  \newcommand{\todo}[1][]{}
  \newcommand{\chI}[0]{}
  \newcommand{\chII}[0]{}
  \newcommand{\chIII}[0]{}
  \newcommand{\chIV}[0]{}
  \newcommand{\ch}[0]{}
\else
  \newcommand{\todo}[1][]{\textbf{\fcolorbox{black}{red}{\color{white}{TODO}}} \underline{$\overline{\hbox{\emph{#1}}}$}}
  \newcommand{\chI}[0]{}
  \newcommand{\chII}[0]{}
  \newcommand{\chIII}[0]{}
  \newcommand{\chIV}[1]{{\textcolor{MidnightBlue}{#1}}}
  \newcommand{\ch}[1]{{\textcolor{BrickRed}{#1}}}
\fi

\begin{document}
%
\title{Guest Editor Introduction:\\ \chII{Recent Advances in DRAM and Flash Memory Architectures}}


\author{%
{Onur Mutlu$^{1,2}$}%
\qquad%
{Saugata Ghose$^{2}$}%
\qquad%
{Rachata Ausavarungnirun$^{2}$}%
}

\affil{\em%
$^{1}$ETH Z{\"u}rich%
\qquad%
$^{2}$Carnegie Mellon University%
}

\maketitle


\chI{Memory and storage systems are a fundamental system
performance, energy, and reliability bottleneck in modern 
systems\chI{~\cite{mutlu.imw13, mutlu.superfri14, mutlu.date17,
cai.procieee17, cai.procieee.arxiv17, cai.bookchapter.arxiv17}}.
This bottleneck is becoming increasingly severe due to 
(1)~the \chII{very} limited latency reductions in memory and storage devices over
the last several years;
(2)~aggressive manufacturing process technology scaling and \chII{other
techniques to improve memory density,}
such as multi-level cell technology, which 
\chII{increase} the \chII{storage capacity} of these devices, but introduce 
more raw bit errors and increase manufacturing process variation;
(3)~limited pin counts in chip packages, which \chII{prevent} system designers
from adding more and/or wider buses to increase bandwidth;
(4)~\chII{overwhelmingly} data-intensive applications, which 
\chII{require high-bandwidth access to very large amounts of data;} and
(5)~the increasing fraction of overall system energy consumed by memory
systems and data movement.
To make matters worse, it is becoming increasingly difficult to continue
scaling these devices to smaller process \chII{technology nodes}, and \chII{even though}
alternative \chII{emerging} memory and storage technologies \chII{can potentially} 
\chIII{alleviate} some of the 
\chII{shortcomings of existing memory and storage technologies, they also}
introduce \emph{new} \chII{shortcomings that were previously absent}.  
Therefore, there is a pressing need to
comprehensively understand and mitigate these bottlenecks in both
existing and emerging \chII{memory and storage} systems and technologies.}

\chI{This issue features extended summaries \chII{and retrospectives} of some of the recent research
done by our group, SAFARI~\cite{safari.website, safari.github},
on (1)~understanding, characterizing, and modeling various critical 
properties of modern DRAM and NAND flash memory,
\chII{the dominant memory and storage technologies, respectively}; and
(2)~several \chII{new} mechanisms we have proposed based
on our observations from these analyses, characterization, and modeling,
to tackle \chII{various key} challenges in memory and storage scaling.
In order to understand the sources of various bottlenecks \chII{of the dominant
memory and storage technologies}, these works
perform rigorous studies of device-level and application-level behavior,
using a combination of detailed simulation and experimental 
characterization of \emph{real} memory and storage devices.}

\chI{The works that perform real device characterization make use of custom
FPGA-based platforms that we build to}
provide us with fine-grained control over the devices.  \chI{We devise}
specific tests that perform a controlled measurement of each phenomenon that
we aim to explore.  \chII{\chIII{Our} experimental} characterizations have often discovered many
unexpected \chIII{types of behavior} in real state-of-the-art devices, and have inspired the
research community to pursue further investigations \chII{(e.g., \chIV{on} the RowHammer
phenomenon~\cite{kim.isca14, mutlu.date17}, DRAM retention 
behavior\ch{~\cite{liu.isca13, khan.sigmetrics14, qureshi.dsn15}},
NAND flash \ch{memory} error patterns\ch{~\cite{cai.date12, cai.date13, cai.procieee17,
cai.procieee.arxiv17, cai.bookchapter.arxiv17, cai.iccd12, cai.hpca15, cai.dsn15,
cai.iccd13}})}.  In order to \ch{aid 
future research}, we have released \chII{much of our experimental} 
characterization data online\chIII{~\cite{safari.tools.website, safari.github}}, and have
open-sourced our DRAM characterization platform, \chII{SoftMC\chIII{~\cite{hassan.hpca17, softmc.github}}}.

\chII{The works that perform application and architectural analyses rely on real
system characterizations and simulation to develop a rigorous understanding of
the bottlenecks and to provide solutions.  
\chIII{Our analyses have shown key scaling bottlenecks, proposed new solutions,
and have inspired the research community to develop further investigations
(e.g., \chIV{on} DRAM refresh\ch{~\cite{liu.isca12, liu.isca13, chang.hpca14}}, DRAM latency
reduction\ch{~\cite{lee.hpca13, lee.hpca15}}, \chIV{the RowHammer phenomenon~\cite{kim.isca14, 
mutlu.date17},} and in-memory data movement and
computation\ch{~\cite{seshadri.micro13, chang.hpca16, seshadri.cal15, seshadri.micro17}}).}
In order to \ch{aid future} research,
we have released our \chIII{flexible and extensible} memory system simulator, Ramulator, as open-source
software~\cite{kim.cal15, ramulator.github}.}

\chI{In each work \chII{that is featured in this issue}, based on our 
observations and analyses from our \chII{experimental} studies 
\chII{of real systems and applications as well as future trends and problems}, we 
propose novel solutions that overcome many of the scaling bottlenecks that
memory and storage systems face.}
For \chII{each of the} works presented \chI{in this special issue}, 
\chII{its corresponding article examines the} work's significance in the
context of modern computer systems, and \chII{discusses} several \chIII{new} research
questions \chIII{and directions} that each work motivates.

We start \chI{with \chIV{five} of our works that explore} new opportunities in \chI{DRAM systems} to
reduce latency and/or energy \chII{consumption}.
\chII{As we mentioned earlier, the \chIII{latency} and energy consumption of DRAM
\chIII{have} not reduced significantly in the last several years.  We find that by
introducing heterogeneity into DRAM architectures, or by taking advantage of the
existing variation within and across DRAM modules, we can develop new mechanisms
that improve DRAM access latency and/or energy efficiency.}

\chI{The first paper in the issue~\cite{tl-dram-safarij} describes} Tiered-Latency DRAM (TL-DRAM), 
which originally appeared in HPCA 2013~\cite{lee.hpca13}.
\chI{This work \chII{(1)~}proposes} a \chII{new} DRAM architecture that can provide us with the performance benefits of
costly reduced-latency DRAM products in a cost-effective manner, by 
isolating a small portion of a DRAM array so that it can behave as
a \chIV{low-latency} DRAM \chII{buffer;
and (2)~exploits the \chIV{low-latency in-DRAM} buffer using various hardware or software mechanisms
to improve overall system performance}.

\chI{The second paper in the issue~\cite{al-dram-safarij} describes} Adaptive-Latency DRAM (AL-DRAM), 
which originally appeared in HPCA 2015~\cite{lee.hpca15}.
\chI{This work experimentally characterizes} 
(1)~the large latency variation across DRAM modules and
(2)~the large timing margins designed to account for worst-case variation and operating
conditions.  \chII{Based on the findings from the characterization, the work}
\chI{proposes} a \chIII{new} mechanism that can
identify and safely reduce the timing margin to speed up DRAM accesses,
\chII{and thus improve overall system performance and energy consumption}.

\chI{The third paper in the issue~\cite{fly-dram-safarij} describes} Flexible-Latency DRAM (FLY-DRAM), 
which originally appeared in SIGMETRICS 2016~\cite{chang.sigmetrics16}.
\chI{This work experimentally characterizes} the latency variation that exists \emph{within} each
DRAM module, \chII{showing that there are regions of fast cells and regions of slow cells
that exist in real DRAM modules.}
\chII{Based on these findings, the work} \chI{proposes} a \chIII{new} mechanism that
identifies regions of fast cells and reduces the latency of DRAM operations to these
regions.

\chI{The fourth paper in the issue~\cite{voltron-safarij} describes} Voltron, 
which originally appeared in SIGMETRICS 2017~\cite{chang.sigmetrics17}.
\chI{This work experimentally characterizes} the relationship between DRAM latency, reliability, and
supply voltage, \chII{showing that these three can be traded off intelligently for
various purposes.} \chII{The work proposes a \chIII{new} mechanism that} \chI{uses} this relationship to 
\chII{dynamically} reduce DRAM energy consumption within
a bounded performance loss target.

\chI{The fifth paper in the issue~\cite{softmc-safarij} describes} SoftMC, 
which originally appeared in HPCA 2017~\cite{hassan.hpca17}.
\chI{This work \chII{describes}} our open-source DRAM characterization infrastructure, and 
\chI{demonstrates} its versatility for use in a wide range of DRAM research topics.
\chI{\chII{SoftMC} is a result of 6+~years of effort, which led to \chII{at least} 11~works at
top conferences, \chII{and we hope it will enable other researchers to
explore the detailed behavior of existing and emerging memory architectures
and develop new mechanisms and memory architectures}.}

Next, we look at \chI{a couple of our works that reduce data movement between the CPU and DRAM,
as this movement consumes (1)~a large fraction of DRAM energy and 
(2)~much of the limited available DRAM bandwidth.}
We find that a large portion of DRAM bandwidth is consumed by the
movement of data between DRAM and the CPU to perform simple operations
such as data copy and initialization.  We can instead take advantage of the underlying
DRAM architecture to efficiently perform these simple operations directly within
DRAM, eliminating the need to move the data \chIII{to/from} the CPU.

\chI{The sixth paper in the issue~\cite{rowclone-safarij} describes} RowClone, 
which originally appeared in MICRO 2013~\cite{seshadri.micro13}.
\chII{Many applications perform data copy and initialization operations,
requiring only simple computation, but these operations require expensive
data movement between the CPU and DRAM.}
\chI{This work proposes} a \chIII{new} DRAM architecture that can \chII{internally}
perform bulk data copy and initialization operations \chII{at very low hardware
cost, \chII{avoiding the costly data movement,} and shows that doing so 
provides 1--2 orders of magnitude speedup and
energy reduction for such operations.}

\chI{The seventh paper in the issue~\cite{lisa-safarij} describes} low-cost interlinked subarrays (LISA), 
which originally appeared in HPCA 2016~\cite{chang.hpca16}.
\chI{This work (1)~builds} a general substrate that facilitates the bulk movement 
of data between two different rows in memory \chII{by improving the
interconnectivity of DRAM arrays}, \chI{and 
(2)~demonstrates} that LISA can be used to efficiently implement a number of
mechanisms, such as bulk \chII{data} copy/initialization, \chII{latency reduction,} and 
fast \chIII{in-DRAM} caching.
\chII{Each of these mechanisms provides significant performance and energy
improvements.}

Finally, we investigate the reliability of NAND flash memory. 
\chII{As NAND flash memory based solid-state drives (SSDs) are now widely-used in 
a large variety of modern systems (e.g., data centers\chIII{~\cite{meza.sigmetrics15,
schroeder.fast16, narayanan.systor16}}, smartphones), there is 
continued demand to increase the density of SSDs while lowering the cost per bit.
While manufacturers have employed several methods (e.g., aggressive 
manufacturing process technology scaling \chIII{and} multi-level cell technology)},
these methods have exacerbated a number of sources of
raw bit errors.  Due to limitations to the number of errors that can be 
corrected by error-correcting codes (ECC), SSDs have a limited lifetime,
after which manufacturers cannot reliably retain data for a minimum guaranteed
time without data loss~\cite{cai.procieee17, cai.procieee.arxiv17, 
cai.bookchapter.arxiv17}.  Over the last decade, as a result of aggressive
density scaling, the typical lifetime of an SSD has dropped by 1--2 orders of 
magnitude, and the various sources of raw bit errors now pose a key scaling 
challenge for storage~\cite{cai.procieee17, cai.procieee.arxiv17, 
cai.bookchapter.arxiv17}.  \chIII{As a sampling of our 7+~years of research into
NAND flash memory reliability, we}
\chI{feature \chIII{three} papers that} design mechanisms
\chI{to} significantly mitigate reliability issues and extend the limited lifetime of
NAND flash memory based devices.

\chI{The eighth paper in the issue~\cite{retentionerr-safarij} describes a new} data retention study \chII{in
NAND flash memory}, 
which originally appeared in HPCA 2015~\cite{cai.hpca15}.
\chI{This work experimentally characterizes} the susceptibility of state-of-the-art NAND flash memory
to \chIII{data} retention errors \chIII{using our FPGA-based flash memory testing
infrastructure~\cite{cai.date12, cai.procieee17, cai.procieee.arxiv17, cai.bookchapter.arxiv17}}, 
and \chI{proposes} \chII{(1)~a \chIII{new} mechanism that mitigates} the impact of
retention errors at runtime, \chII{which increases the lifetime of the SSD;} and 
\chII{(2)~a \chIII{new} mechanism that exploits} retention behavior to recover data 
in the event of data loss, \chIII{thereby improving SSD robustness}.

\chI{The ninth paper in the issue~\cite{read-dist-safarij} describes a new} read disturb study \chII{in
NAND flash memory}, 
which originally appeared in DSN 2015~\cite{cai.dsn15}.
\chI{This work experimentally characterizes} read disturb errors in NAND flash memory, where a
read operation introduces errors in unread parts of the memory. 
\chII{Based on the characterization, the work}
\chI{proposes} \chII{(1)~a \chIII{new} mechanism that mitigates read disturb errors,
thereby improving the SSD lifetime; and
(2)~a \chIII{new} mechanism that exploits read disturb behavior to recover data in the 
event of data loss, \chIII{thereby improving SSD robustness}.}

\chI{The last paper in the issue~\cite{char-safarij} describes a new study on two-step programming
in NAND flash memory}, 
which originally appeared in HPCA 2017~\cite{cai.hpca17}.
\chI{This work demonstrates} that the programming algorithm used in many state-of-the-art 
NAND flash memory devices can introduce previously-unknown data
vulnerabilities, which can be exploited by malicious applications to perform security attacks.
\chI{The work proposes} three mechanisms to eliminate or mitigate these vulnerabilities,
\chII{thereby improving both reliability and security}.

Throughout all of these works, we find that by understanding and taking advantage of
the \chII{behavior and architecture} of memory and storage devices \chIII{and
appropriately modifying them at low cost and low overhead}, we can successfully 
\chII{mitigate} many of the scalability challenges in memory and storage devices.
\chIII{Even though the works presented are described in the context of DRAM and
NAND flash memory, the two dominant memory and storage technologies of today,
we believe many of the basic ideas and concepts can be applied or \chIV{adapted} to
emerging memory technologies\chIV{~\cite{meza.weed13}},
e.g., phase-change memory~\cite{lee.isca09, qureshi.isca09, wong.procieee10, 
lee.ieeemicro10, zhou.isca09, lee.cacm10, yoon.taco14},
STT-MRAM~\cite{naeimi.itj13, kultursay.ispass13, guo.isca10},
and memristors/RRAM~\cite{wong.procieee12, chua.tct71, strukov.nature08}.}
We hope that \chIII{the works featured in this special issue} inspire readers to explore \chII{the presented} challenges, and
to develop new solutions that can enable high-performance, \chI{low-energy,
low-latency,} high-reliability \chII{memory and storage} systems, \chII{and
thus the computing systems, of} the future.

\section*{Acknowledgments}

The works featured in this issue, along with our related works that
we reference in each \chII{featured work}, are a
result of the research done together with many students and
collaborators over the course of the past 10+ years, whose
contributions we acknowledge.
In particular, we acknowledge and appreciate the dedicated effort 
of current and former students and postdocs in our research group,
\chI{SAFARI~\cite{safari.website, safari.github}},
who contributed to the \chI{ten} featured works, 
including
Yu Cai,
Kevin Chang,
Chris Fallin,
Hasan Hassan,
Kevin Hsieh,
Ben Jaiyen,
Abhijith Kashyap,
Samira Khan, 
Yoongu Kim, 
Tianshi Li, 
Jamie Liu, 
Donghyuk Lee, 
Yixin Luo, 
Justin Meza, 
Gennady Pekhimenko, 
Vivek Seshadri, 
Lavanya Subramanian, 
Nandita Vijaykumar, and
Abdullah Giray Ya{\u g}l{\i}k{\c c}{\i}.

Aside from our featured
works and other referenced papers from our group, where a wealth
of information on modern memory and storage systems can be found, at least four
Ph.D.\ dissertations have shaped the works that we feature \chI{in this special issue}:
\begin{itemize}[leftmargin=1.6em]
\item Yu Cai's thesis
entitled ``NAND Flash Memory: Characterization, Analysis, Modeling and
Mechanisms''~\cite{cai.thesis12},

\item Donghyuk Lee's thesis entitled
``Reducing DRAM Latency at Low Cost by Exploiting
Heterogeneity''~\cite{lee.thesis16},

\item Vivek Seshadri's
thesis entitled ``Simple DRAM and Virtual Memory Abstractions to
Enable Highly Efficient Memory Subsystems''~\cite{seshadri.thesis16}, and

\item Kevin Chang's thesis entitled ``Understanding and
Improving the Latency of DRAM-Based Memory
Systems''~\cite{chang.thesis17}.
\end{itemize}

We also acknowledge various funding agencies (the National Science
Foundation, the Semiconductor Research Corporation, the Intel Science and
Technology Center on Cloud Computing, CyLab, the CMU Data Storage
Systems Center, and the NIH) and
industrial partners (AMD, Facebook, Google, HP Labs, Huawei, IBM,
Intel, Microsoft, NVIDIA, Oracle, Qualcomm, Rambus, Samsung, Seagate,
VMware), and ETH Z{\"u}rich, who have supported the featured works in
this issue and other related work \chII{in our research group} generously 
over the years.

{
\bibliographystyle{IEEEtranS}
\bibliography{refs}
}

\end{document}